\documentclass[10pt,letterpaper,twocolumn]{revtex4-1} 

\usepackage{amsmath}
\usepackage{graphicx}

\newcommand{\unit}[1]{\ensuremath{\,\mathrm{#1}}}

\newcommand{\hypref}[2]{\hyperref[#2]{{#1}~\ref{#2}}}

\graphicspath{{images/}}

\begin{document}


\title{Sub-kHz lasing of a CaF$_2$ Whispering Gallery Mode Resonator\\Stabilized Fiber Ring Laser}


\author{M. C. Collodo$^{1,2}$, F. Sedlmeir$^{1,2,5}$, B. Sprenger$^3$, S. Svitlov$^2$, L. J. Wang$^4$, and H. G. L. Schwefel$^{1,2,*}$}

\address{$^1$Max~Planck~Institute~for~the~Science~of~Light, G.-Scharowsky-Str.~1/Bau~24, 91058 Erlangen, Germany\\
$^2$Institute of Optics, Information and Photonics, University of Erlangen-Nuremberg, G.-Scharowsky-Str.~1/Bau~24, 91058~Erlangen, Germany\\
$^3$Humboldt-Universit\"at zu Berlin, 
Institut f\"ur Physik, AG Nanooptik,
Newtonstra\ss e 15, 12489 Berlin, Germany\\
$^4$Physics~Department~and~Joint~Inst.~of~Measurement~Science~(JMI), Tsinghua~University, Beijing, 100084~China\\
$^5$SAOT, School in Advanced Optical Technologies, Paul-Gordan-Stra{\ss}e 6, 91052 Erlangen, Germany\\
$^*$Corresponding author: +49 9131 6877-134, Harald.Schwefel@mpl.mpg.de
}

\begin{abstract}
We utilize a high quality calcium fluoride whispering-gallery-mode resonator to stabilize a simple erbium doped fiber ring laser with an emission frequency of 196\unit{THz} (wavelenght 1530\unit{nm}) to a linewidth below 650\unit{Hz}. This corresponds to a relative stability of $3.3\times10^{-12}$ over 16\unit{\mu s}. In order to characterize the linewidth we use two identical self-built lasers and a commercial laser to determine the individual lasing linewidth via the three-cornered hat method. 
\end{abstract}

\maketitle


\noindent
Compact and stable light sources are in high demand in metrology\cite{lea_limits_2007} and biochemical sensing\cite{baaske_optical_2012} to mention just two predominant fields. Optical resonators are at the heart of both applications. Whispering gallery mode (WGM) resonators are dielectric cavities that confine light due to total internal reflection at their dielectric interface\cite{vahala_optical_2003}. Their quality factor ($Q$) is mainly limited by surface scattering and by material absorption. For highly transparent materials such as calcium fluoride (CaF$_2$) quality factors up to $10^{11}$\cite{grudinin_ultra_2006, savchenkov_optical_2007} have been shown. Operability exists throughout the whole transparency window of the host material, in case of calcium fluoride from 150\unit{nm} to 10\unit{\mu m}. With their resulting very narrow linewidth, these resonators serve as excellent optical frequency filters\cite{matsko_whispering-gallery-mode_2007} and are suitable to enhance the lasing modes of a conventional primary lasing module\cite{liang_whispering-gallery-mode-resonator-based_2010}.

Before reaching the fundamental thermal noise floor limit\cite{alnis_thermal-noise-limited_2011, numata_thermal-noise_2004, chijioke_thermal_2011, matsko_whispering-gallery-mode_2007}, the stability of a reference cavity is mainly determined by its deformation due to mechanical vibration or thermal effects, whose influence scales with the cavity's dimensions \cite{sprenger_caf2_2010}. Therefore, due to their compact sizes, WGM resonators are eminently suitable as a frequency reference.

In this Letter we report the setup and characterization of a free running fiber ring laser providing lasing linewidths below 1\unit{kHz}. This is achieved by resonantly filtering the broad emission spectrum of an erbium doped fiber via the narrow-linewidth modes of a WGM resonator. Only these narrow modes can pass the resonator and achieve gain in the following round trip. This establishes a narrow linewidth lasing behavior. 

In our experiment we observed a suppression of the lasing linewidth to sub-kilohertz in comparison with the cold cavity linewidth of the resonator (sub-megahertz). This can be equivalently described by an improvement of the resonator's $Q$ by a factor of $10^3$ due to active lasing. In order to verify these results, we modeled our ring laser setup analytically following and extending the approach by Wang \cite{wang_causal_2002}. 
We modeled the filtering mechanism due to the WGM resonator and implemented this model in an iterative numerical simulation, taking into account gain saturation. The experimentally observed characteristics could be reproduced by appropriate choice of parameters, the most crucial being the fiber cavity's and the WGM resonator's $Q$ factors and the saturated intracavity power. The laser's emission spectrum narrows with increasing intracavity power. This outcome agrees well with a fully analytic approach by Eichhorn et al. \cite{Eichhorn12}. Our simulation does not cover further aspects regarding the WGM resonator's instability due to an increased lasing power, and is therefore not able to determine the optimal intracavity power.

\begin{figure}[tb]
\centering
\includegraphics[width=8.4cm]{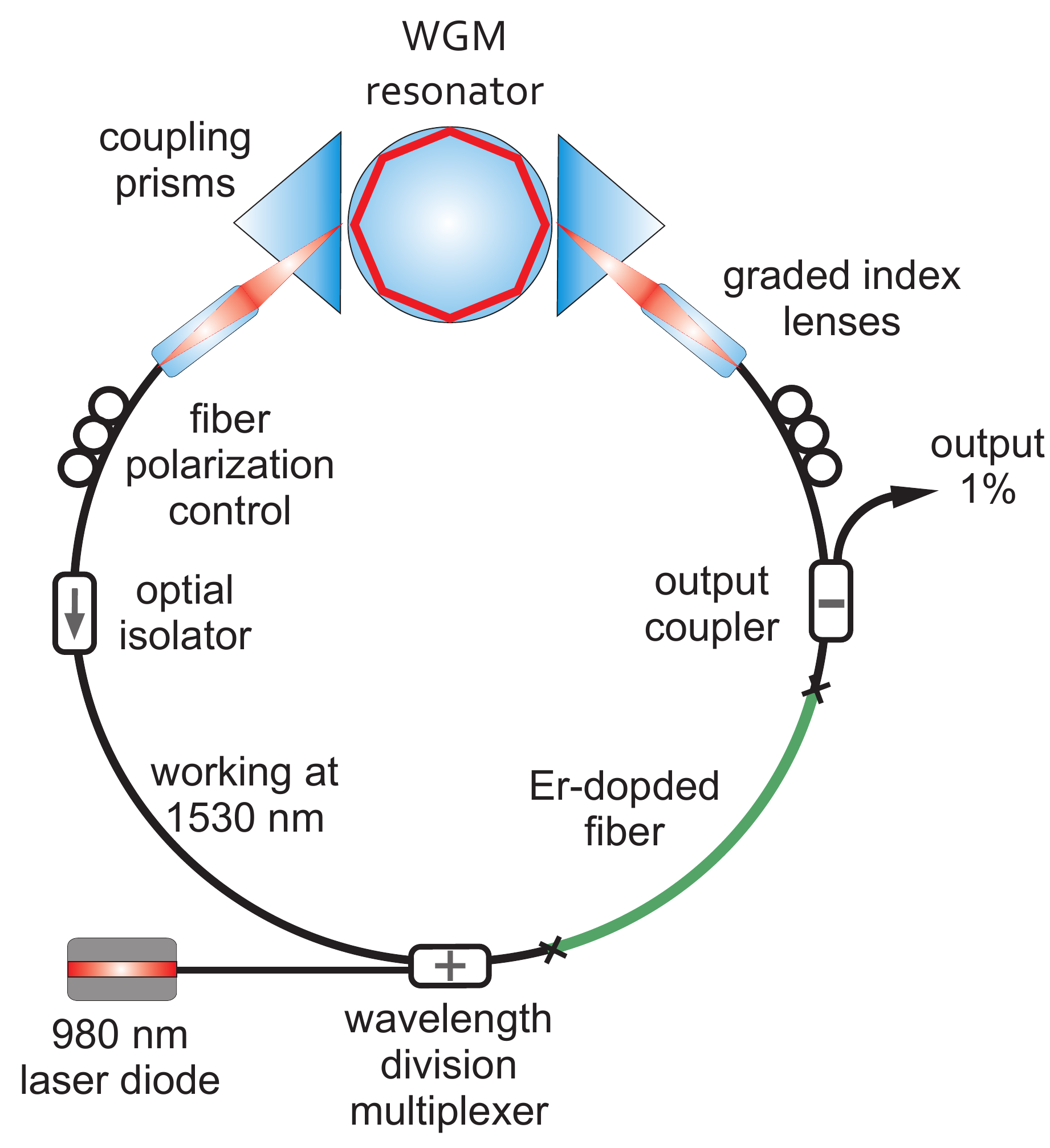}
\caption{(Color online) Sketch of one of the whispering gallery mode resonator filtered lasers. A WGM resonator is used to filter the emission spectrum of a conventional telecom fiber ring laser. The active medium is an erbium doped fiber with an emission spectrum in the telecommunication C-band.
Passive filtering of the fiber loop lasing modes is provided by a WGM resonator.
Single mode lasing is obtained without further active stabilization techniques.}
\label{fig:setup}
\end{figure}

\textsl{Experimental setup. }
We set up a conventional fiber ring laser using an erbium doped fiber with a broad band emission spectrum in the telecommunication C-band wavelength regime ($\sim$1530\unit{nm}). It is pumped by a 980\unit{nm} laser diode with 200\unit{mW} output power through a wavelength-division-multiplexer. 
Light is coupled out via a 99/1 fiber coupler and prevented from clockwise circulation by an optical Faraday isolator.
By inserting the WGM resonator into the fiber loop the final whispering gallery laser (WGL) is set up (figure \ref{fig:setup}). 
We fabricate the WGM resonators on a home built diamond lathe. Mono-crystalline CaF$_2$ is cut into disks with an optimized surface curvature via diamond turning. The disks are 4\unit{mm} in diameter. An optimal surface quality is achieved via polishing with grain sizes down to 50\unit{nm}. Loaded cold cavity $Q$ factors of our millimeter sized resonators measure a few $10^8$. Evanescent coupling through the polarization dependent WGMs can be achieved via a pair of piezo controlled coupling prisms (SF11 glass). Prism coupling allows for a more rigid construction, which is less influenceable by mechanical vibrations. In comparison, previous approaches utilized tapered or angle polished fibers for the resonator coupling \cite{sprenger_whispering-gallery-mode-resonator-stabilized_2009}. For the coupling into and out of the fiber loop a pair of gradient index (GRIN) lenses is used, where the numerical aperture and the focal point are adjustable. The coupling efficiency from the fiber loop transmitted through the WGM resonator was approximately 20\%. For optimal coupling to distinct whispering gallery modes a fiber polarisation controller is necessary. Laser output power is in the range of tens of microwatts.
This solely passively stabilized lasing systems provides a straightforward setup design, featuring easy assembly and tight packaging.

As our main task will be the linewidth measurement, the tunability of the lasing frequency is paramount. With a broadband ($1\unit{nm}$ full width at half maximum) optical bandpass filter (not depicted) a coarse tuning of the lasing mode's wavelength over the whole emission spectrum of the erbium doped fiber in steps of the WGM resonator's free spectral range ($\sim$20\unit{GHz}) is possible. Further fine tuning can be achieved via temperature control of the resonator.

\begin{figure}[tb]
\flushleft
{\raisebox{-0.5cm}{(a)}}
\includegraphics[width=8.4cm]{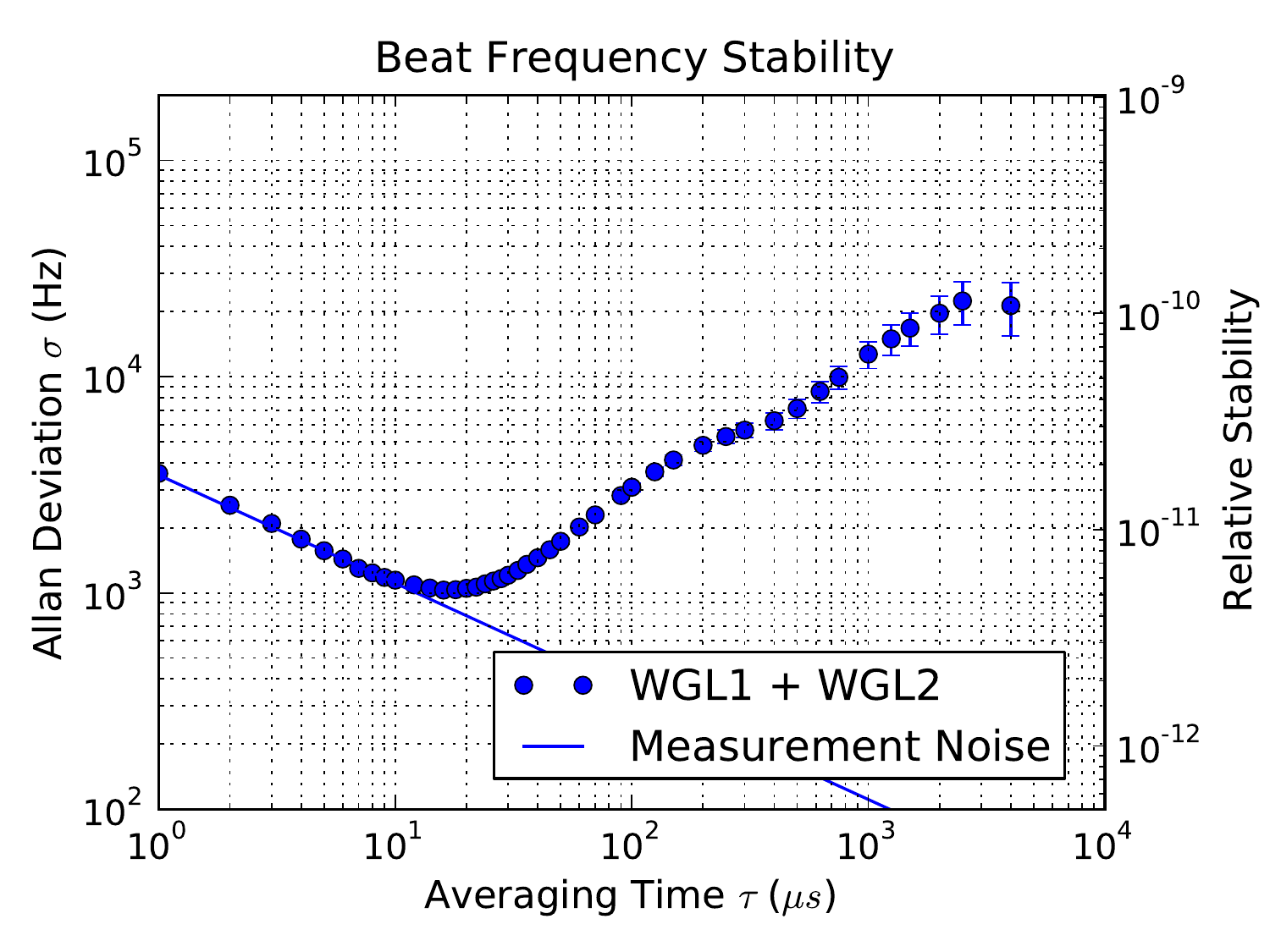}
{\raisebox{-0.5cm}{(b)}}
\includegraphics[width=8.4cm]{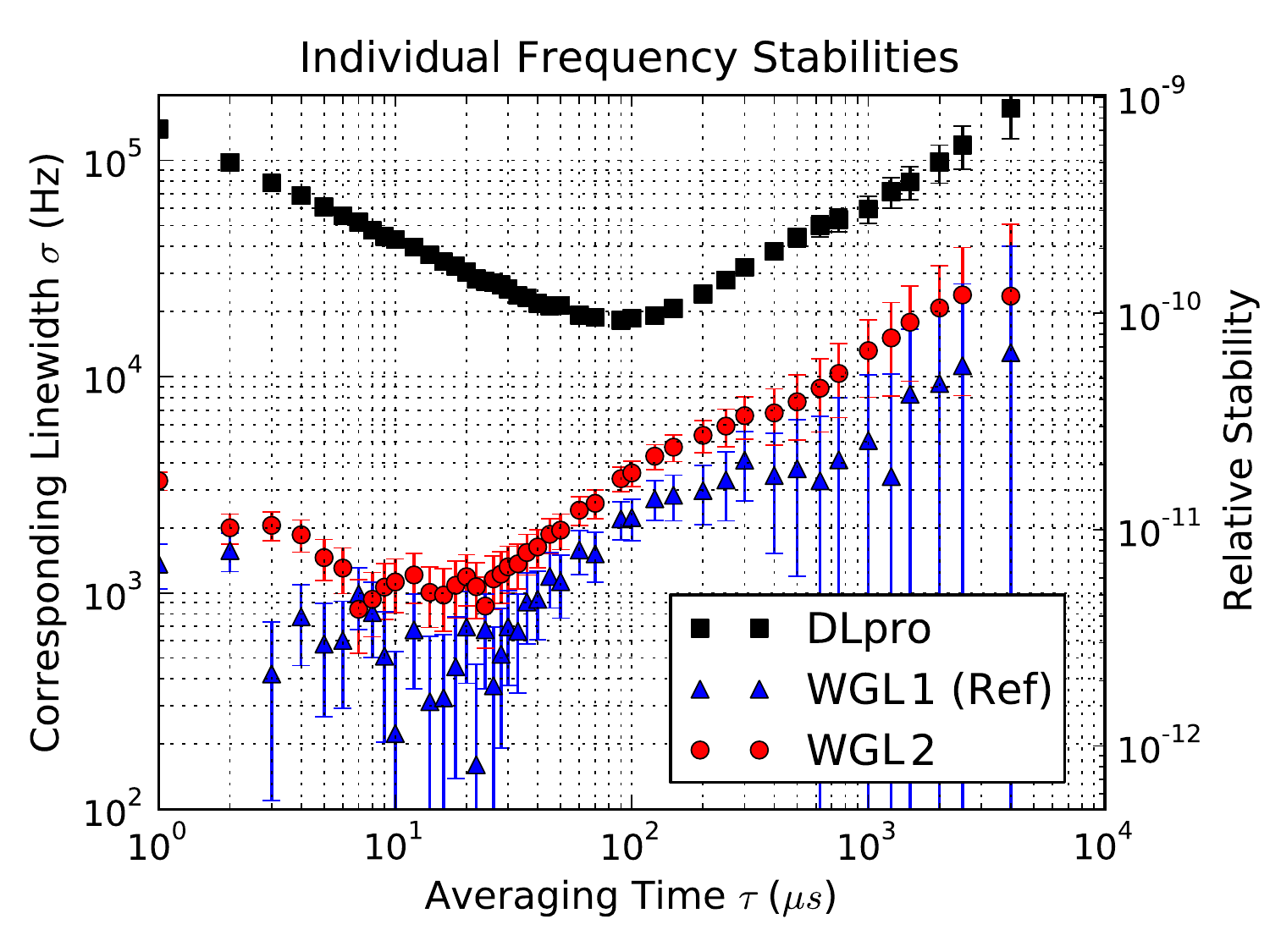}
\caption{(Color online) Allan Deviation values (corresponding to lasing linewidth in \unit{Hz}) and relative stabilities (corresponding to lasing $Q$ factor) for the whispering gallery lasers (WGL). (a) Direct evaluation of the beat note signal between WGL1 and WGL2 reports the combination of both noise sources. (b) Individual noise components were obtained via the three-cornered hat method, a possible correlation was taken into account.}
\label{fig:allan_3ch}
\end{figure}
\textsl{Measurement procedure. }
As the frequency stability of the individual lasers cannot be measured directly, two identical systems were built. The beat note generated by mixing the emission specta of the two lasers allows us to reconstruct the frequency stability of the combined system. We recorded 50\unit{ms} (sampling rate is 200\hspace{0.5ex}megasamples/second, with 8 bit vertical resolution) of the beat note traces of the two approximately 10\unit{MHz} detuned WGLs. For the computation of the beat note's frequency contiguous, non overlapping basic time intervals of 1$\unit{\mu s}$ length are chosen from the trace in order to obtain a frequency stream.

This basic time interval is chosen such that it covers at least ten periods of the beat note signal. We perform a nonlinear least-square fit of a sine function over these separate time domain traces. The only fitting parameters in the used model are frequency, amplitude, offset and phase, which are all assumed to be constant over the basic 1$\unit{\mu s}$ fit interval.
We prefer this method in comparison to a conventional frequency counter because of the higher flexibility and robustness in case of a low signal to noise ratio. Furthermore, we avoid ambiguity in the interpretation of the resulting values and we are free to compute different types of variances\cite{rubiola_measurement_2005}.

In order to determine the frequency stability of our lasers with respect to the averaging time the Allan Deviation\cite{allan_standard_1988,barnes_characterization_1971} 
\begin{equation}
\sigma^2 = \frac{1}{2(M-1)} \sum_{i=1}^{M-1} \left(f_i - f_{i+1} \right)^2
\end{equation}
is used, where $\sigma$ is the Allan Deviation of the lasing frequency, $f_i$ denotes the value of the lasing frequency referring to the $i$-th basic time interval and $M$ is the total amount of basic time intervals.
Pre-averaging of the frequency stream over multiple basic time intervals allows us to compute Allan Deviations for different averaging time scales.

\textsl{Results. }
The resulting Allan Deviation values are shown in figure \ref{fig:allan_3ch} (a). They directly correspond to the lasing linewidth. The relative stabilities comparing the laser's linewidth to its emission frequency of 196\unit{THz} are also presented, they correspond to the inverse lasing $Q$ factor.
For the deviation values of the beat note trace evaluation a minimum of 1056\unit{Hz} is reached for an averaging time of 18\unit{\mu s}. Since the individual WGM laser setups are slightly different we are reporting an upper limit for the more stable WGL. This can be estimated by 1056/$\sqrt{2}$\unit{Hz} = 750\unit{Hz}. Thereby we assume no negative correlation in the lasing behavior.
The graph (figure~\ref{fig:allan_3ch} (a)) reveals a $\tau^{-1/2}$-slope in the short timescale regime (less than 18\unit{\mu s}), before the optimal averaging time is attained. This slope can be associated with a white frequency noise behavior \cite{allan_standard_1988}. After the optimal averaging time a drift to larger Allan Deviation values is predominant. The timescale suggests that temperature fluctuations alone cannot be responsible. Also the curve's slope does not fit to the therewith related random frequency walk ($\tau^{1/2}$). A directed frequency shift due to heating of the modal volume seems more plausible here.
The measurement noise curve in figure~\ref{fig:allan_3ch}~(a)  is a measure for the quality of the signal and of the measurement procedure as a whole. It combines the frequency errors computed from the least square residuals and thus reflects the signal to noise ratio of the acquired beat note, time jitter and quantization errors of the oscilloscope, errors due to the method of frequency calculation and the short term (less than 1\unit{\mu s}) instability of the lasers, which is ignored by the fit model.

In order to extract the individual frequency stabilities we add a third lasing system, namely a commercial Toptica DLpro design laser and perform a three-cornered hat measurement \cite{gray_method_1974}. This is done by recording the beat notes from the three possible combinations of laser pairs simultaneously and solving for the single laser variances
\begin{equation}
2\sigma^2_{\text{\tiny{WGL1}}} = \sigma^2_{\text{\tiny{WGL1+WGL2}}} + \sigma^2_{\text{\tiny{WGL1+DLpro}}} - \sigma^2_{\text{\tiny{WGL2+DLpro}}}
\end{equation}
(and permutations of this equation).
This holds only if no correlation in the lasing frequency characteristics is predominant. For a more incontestable approach taking into account possible correlations, a correlation removal algorithm by Premoli et al.\ \cite{premoli_revisited_1993} has been used with WGL1 as a reference (choice of reference is not significant). The individual frequency stabilities, revised in this manner, are shown in figure \ref{fig:allan_3ch}~(b).

The more stable WGM resonator laser reaches a relative stability of $\left(1.67 \pm 1.60 \right) \times 10^{-12}$ for an averaging time of 16\unit{\mu s}. This corresponds to a lasing linewidth of $\left(328 \pm 314 \right) \unit{Hz}$ at the laser's emission frequency of 196\unit{THz}. In a conservative estimate we report a relative stability of $3.3 \times 10^{-12}$ and a linewidth of 650\unit{Hz}, respectively. These values agree well with the results of the prior estimation using the combined frequency stability directly.

\textsl{Conclusions. }
To summarize, we demonstrated a sub-kilohertz linewidth lasing behavior in a solely passively stabilized  erbium doped fiber ring laser. The stabilization arises through filtering via high-$Q$ modes of a crystalline calcium fluoride whispering gallery mode resonator. Our evaluation method for the lasing stability is based on the analysis of the digitized time domain beat note traces and avoids the standard frequency counter approach. The experimentally observed linewidth enhancement during the lasing process to finite lasing $Q$ factors is confirmed theoretically. Thus, the final lasing linewidth can be influenced either by the cold cavity $Q$ factor of the filtering resonator or by an increased circulating intracavity power.

The authors would like to thank Dmitry V. Strekalov and Josef U. F\"urst for stimulating discussions and Gerd~Leuchs for his support.
\bibliographystyle{osajnl}

\end{document}